\documentclass{aa}
\usepackage{graphics}
\usepackage{amssymb}
\usepackage{times}

\newcommand{\de}{$^{\circ}$}

\begin{document}

\thesaurus{11(11.04.2; 11.09.1 Sagittarius; 11.12.1; 12.07.1)}

\title{The Sagittarius dwarf galaxy as a microlensing target}

\author{P. Cseresnjes \inst{1,2} \and C. Alard \inst{1,2,3}}

\offprints{patrick.cseresnjes@obspm.fr}

\institute{DASGAL, Observatoire de Paris, 61 Avenue de l'Observatoire, F-75014 Paris
      \and Centre d'Analyse des Images - INSU
      \and Institut d'Astrophysique de Paris, 98 bis Boulevard Arago, F-75014 Paris}

\date{Received .../ Accepted ...}

\authorrunning{P. Cseresnjes \& C. Alard}

\maketitle

\begin{abstract}
We estimate the optical depth, time-scale distribution and fraction of microlensing events originating from sources in the Sagittarius dwarf galaxy (Sgr) lensed by deflectors in the Milky Way. These events have a time-scale longer by a factor $\sim$1.3 than the MW/MW events and occur mainly on sources fainter than V$\sim$21 mag below Sgr's turn off. The fraction of events involving a source in Sgr depends on the location and extinction of the field and on the limiting magnitude of the survey. The contribution of the MW/Sgr events is negligible ($\lesssim$1$\%$) at very low latitudes ($|b|\lesssim$2\de) but increases continuously towards higher $|b|$ and becomes dominant near the highest density region of the dwarf galaxy. Sgr is present within the fields of current microlensing surveys and any optical depth map inferred from observations will become biased by the presence of Sgr towards higher $|b|$ where the contribution of MW/Sgr events is significant. Systematic spectroscopic measurements on the sources of all the microlensing events may allow detection of this kind of event for which the degeneracy on the lens mass can be significantly reduced.
\keywords{Galaxies: dwarf - Galaxies: individual: Sagittarius dwarf - Local group - Gravitational lensing}
\end{abstract}
%
%
\section{Introduction}
\indent More than $\sim$350 microlensing events detected towards the Galactic Centre have been published to date by several collaborations: MACHO (Alcock et al. \cite{macho1}; Alcock et al. \cite{macho2}; Alcock et al. \cite{macho3}, hereafter A00), OGLE (Udalski et al. \cite{ogle1};  Udalski et al. \cite{ogle2}) and DUO (Alard \& Guibert \cite{duo}). The resulting optical depth is $\gtrsim 3\times 10^{-6}$ in Baade's window, a value that is higher than the theoretical expectation. Several authors have tried to reconcile the observations with the theoretical calculation (Han \& Chang \cite{h98}, Peale \cite{peale}) but too many uncertainties remain in the Bulge structure and kinematics to allow a reliable determination of the lens population. The bulk of the observed events is believed to be caused by self-lensing within a bar making a small angle relative to the line of sight (Paczy\'nski et al. \cite{pac2}, Zhao \& Mao \cite{zm}), but a substantial number of events (10$\%$-20$\%$) may be due to lensing involving Disk stars (Mollerach \& Roulet \cite{mr}, Nair \& Miralda-Escud\'e \cite{nair}). \\
\indent It is, however, also possible for some microlensing events detected towards the Galactic Centre to be caused by sources in the Sagittarius dwarf galaxy (Sgr) lensed by Galactic stars (hereafter MW/Sgr events). Sgr is located $\sim$16 kpc behind the Galactic Centre (Ibata et al. \cite{iwgis}). Its highest surface density region is located at (l=5.6\de, b=-14.0\de) and it is oriented roughly perpendicular to the Galactic Plane so that its northern extension (in galactic coordinates) is present within some fields of microlensing experiments towards the Bulge (Alard \cite{a96}, Alcock et al. \cite{macho_rr}). The Sun/Bulge/Sgr configuration favors a high optical depth for member stars of this dwarf galaxy because of the large value of the distance ratio. Recently, Cseresnjes, Alard \& Guibert (\cite{cag}; hereafter CAG) published a density map of Sgr based on the RR Lyrae distribution between b=-14.0\de and b=-4.0\de and it is now possible to tackle this question quantitatively. \\
\indent In this paper we estimate the fraction of events involving a source in Sgr using a simple galactic model. Sect. 2 describes the models we use for our study. In Sect. 3 we estimate the optical depth, event rate, fraction of events involving a source in Sgr and we consider alternatives to our model to estimate the uncertainties. In Sect. 4 we apply our calculation to current microlensing surveys. Sect. 5 is devoted to a brief discussion about the scientific interests of MW/Sgr events, and is concluded in Sect. 6.
%
%
\section{Models}
\subsection{Milky Way}
\subsubsection{Bulge}
\indent The Bulge is modeled by a bar with a total mass of 2.0 10$^{10}$ M$_{\odot}$ and a triaxial distribution corresponding to the best fit model (G2) to the COBE/DIRBE map (Dwek et al. \cite{dwek})
\begin{equation}
 \rho_{b}=2.30 \ \rm{exp}\bigg[\frac{-\rm{s}^{2}}{2}\bigg] \ (\rm{M}_{\odot}.\rm{pc}^{-3})
\end{equation}
 with
\begin{equation}
 \rm{s}^{4}=\bigg [\bigg(\frac{x'}{1.58\,\rm{kpc}}\bigg)^{2}+\bigg(\frac{y'}{0.62\,\rm{kpc}}\bigg)^{2}\bigg]^{2}+\bigg(\frac{z'}{0.43\,\rm{kpc}}\bigg)^{4}
 \label{s4}
\end{equation}
where $x'$ and $y'$ are the distances along respectively the major axis and minor axis of the Bar in the Disk plane, $z'$ is the distance along the minor axis of the bar perpendicular to the Disk plane. We take the orientation of the bar parallel to the direction (l,b)=(-20$^{\circ}$,0$^{\circ}$). 
\begin{figure}
 \resizebox{\hsize}{!}{\includegraphics{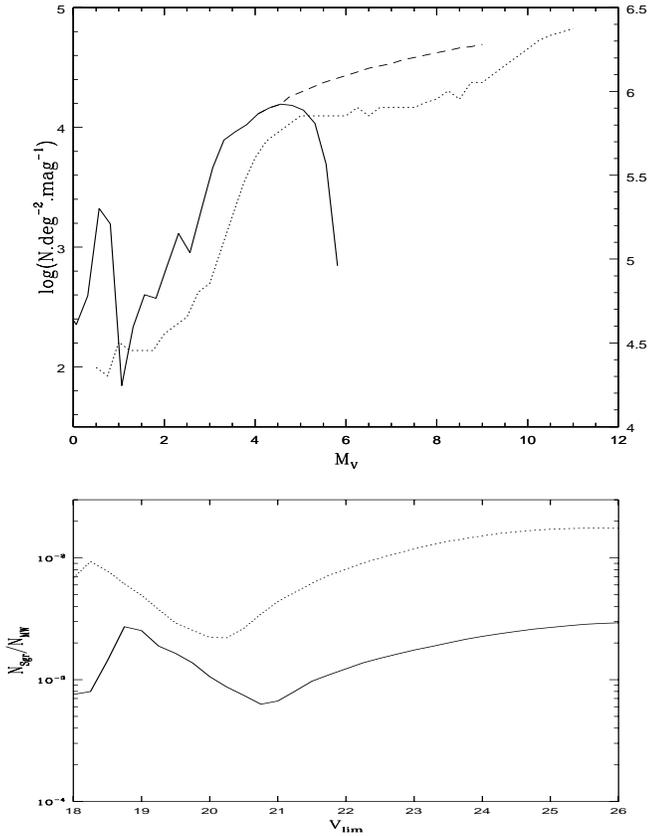}}
 \caption{Upper panel: Luminosity functions of Sgr (full line) and of the Bulge (dotted line). The dashed line is the extrapolation of the luminosity function of Sgr. The Bulge luminosity function is normalized for a field towards Baade's window (right ordinate). The Sgr luminosity function is normalized for a RRab density of 80 deg$^{-2}$ (left ordinate). Lower panel: Fraction of member stars of Sgr as a function of the limiting magnitude. The full (resp. dotted) line corresponds to a field centred on (l,b)=(2.7\de,-3.3\de) (resp. (5.0\de,-5.5\de)).}
 \label{lf}
\end{figure}
 The velocity dispersion is taken from Han \& Gould (\cite{hg}) who calculated $(\sigma_{x'},\sigma_{y'},\sigma_{z'})$=(115.7, 90.0, 78.6) km.s$^{-1}$ from the tensor virial theorem. The distance to the Galactic Centre is set to 8 kpc.\\
\indent The galactic surface stellar density in the direction (l,b) is estimated from the Bulge luminosity function (Fig. \ref{lf}) obtained with the \textit{Hubble Space Telescope} (HST) towards Baade's window (Holtzman et al. \cite{holtz}):
\begin{equation}
 \Sigma=\bigg(\frac{\int_{(l,b)}\rho(x)\,x^{2}\,{\rm d}x}{\int_{BW}\rho(x)\,x^{2}\,{\rm d}x}\bigg) \times \int_{L_{min}}^{L_{max}}\Phi_{BW}(L)\,{\rm d}L
\end{equation}
where $L_{min}$ corresponds to the limiting magnitude. The extinction is taken from Schlegel et al. (\cite{schlegel}) except at low galactic latitudes where the extinction is inferred from the colors of RR Lyrae at minimum light in the MACHO field when available (Alcock et al. \cite{macho_rr}).
\subsubsection{Disk}
\indent For the Disk, we adopt the model of Gould, Bahcall \& Flynn (\cite{gould}) who fitted the M dwarf spatial distribution obtained with the HST
\begin{eqnarray}
 \rho_{d}=0.63 \ \rm{exp}\bigg[-\frac{R}{3.5\,\rm{kpc}}\bigg] \times \bigg\{0.2 \ \rm{exp}\bigg[-\frac{|z|}{0.64\,\rm{kpc}}\bigg] \nonumber \\
 +0.8 \ \rm{sech}^{2}\bigg[\frac{z}{0.32\, \rm{kpc}}\bigg]\bigg\} \ (\rm{M}_{\odot}.\rm{pc}^{-3}).
\end{eqnarray}
The density distribution is normalized in order to get a local surface density of 50 M$_{\odot}$.pc$^{-2}$, consistent with current dynamical estimates (Flynn \& Fuchs \cite{ff}). Since the luminosity function of the Disk at high $|z|$ (where most of the Disk sources are located) is poorly constrained, we take the same luminosity function as for the Bulge. We adopt v$_{c}$=220 km.s$^{-1}$ for the circular velocity of the local standard of rest and ($\sigma_{l},\sigma_{b}$)=(20.0,16.0) km.s$^{-1}$ for the velocity dispersion of the Disk. Finally, for the peculiar motion of the Sun, we take v$_{pec}$=16.5 km.s$^{-1}$ directed towards (l=53\de,b=25\de). 
\subsection{Sgr dwarf galaxy}
\indent The Sagittarius dwarf galaxy is modeled by the surface density map obtained by CAG from RR Lyrae counts. The RRab density is converted into a stellar density through the luminosity function published by Mateo et al. (\cite{muskkk}). The size of their field is 0.063 deg$^{2}$ and it contains 5 RRab, corresponding to a RRab density of $\sim$80$\pm$35 deg$^{-2}$ where the uncertainty represents the 1$\sigma$ Poissonian noise. For comparison, a symmetric field  relative to the minor axis of Sgr contains $\sim$60 RRab stars in the map published by CAG. The Sgr stellar density can thus be estimated by
\begin{equation}
 \Sigma_{Sgr}=\frac{N_{RRab}}{80}\times \int_{L_{min}}^{L_{max}} \Phi_{Sgr}(L)\,dL
\end{equation}
where $N_{RRab}$ is the RRab density in Sgr. The luminosity function of Mateo et al. is nearly complete until M$_{V}\sim$4.5, which is $\sim$1 mag below the Turn Off of Sgr (Marconi et al. \cite{marconi}). We extrapolate this function to M$_{V}$=9 by taking the mean slope of this function between M$_{V}$=3.5 and M$_{V}$=4.5 (Fig \ref{lf}). \\
\indent The kinematical data of Sgr are taken from Ibata et al (\cite{iwgis}) who found (U,V,W)=(232,0:,194) km.s$^{-1}$. Due to the lack of an adequate reference frame, the V component is at the moment poorly constrained. However, the elongation of Sgr which is roughly perpendicular to the Galactic plane suggests a low value for V. The radial velocity dispersion of stars in Sgr is $\sigma_{r}$=11.4 km.s$^{-1}$. We will assume that the velocity dispersion in Sgr is isotropic although this is probably not true due to its elongated shape. However, this has little consequence in our study since the overall dispersion in the relative velocity is highly dominated by the dispersion of the galactic lenses.
%
%
\section{Optical depth and event rate}
\subsection{Optical depth}
\begin{table}
 \caption{Optical depths ($\times$10$^{6}$), event rates ($\times$10$^{6}$ stars $\times$ 1 year) and time-scales (days). The rates and time-scales are calculated for a single lens mass population of 1 M$_{\odot}$.}
 \begin{tabular*}{8.5cm}{l @{\extracolsep{\fill}} c @{  } c @{  } c @{  } c @{  } c @{ } c @{ } c @{ } c}
  \multicolumn{2}{c}{ }&\multicolumn{3}{c}{MW/MW}&\multicolumn{1}{c}{ } & \multicolumn{3}{c}{MW/Sgr} \\
  \cline{3-5} \cline{7-9}
  V$_{\rm lim}$  &  (l,b)                & $\tau_{\rm{MW}}$   &  $\Gamma$ & $\langle\rm{t}_{E}\rangle$ & & $\tau_{\rm{sgr}}$ & $\Gamma$ & $\langle\rm{t}_{E}\rangle$ \\
  \hline
  21.0           &  (2.7\de,-3.3\de)     & 2.2 & 12.2                & 41.3              &       & 17.6 & 68.4                & 58.3 \\
                 &  (5.0\de,-5.5\de)     & 1.2 & 5.4                 & 47.9              &       & 7.7  & 28.5                & 61.5 \\
                 &  (6.0\de,-9.0\de)     & 0.3 & 1.5                 & 48.3              &       & 1.7  & 6.3                 & 62.6 \\
  \hline
  23.0           &  (2.7\de,-3.3\de)     & 2.7 & 14.2                & 42.9              &       & 17.6 & 68.9                & 58.2 \\
                 &  (5.0\de,-5.5\de)     & 1.3 & 5.9                 & 49.2              &       & 7.7  & 28.7                & 61.5 \\
                 &  (6.0\de,-9.0\de)     & 0.4 & 1.6                 & 49.3              &       & 1.7  & 6.3                 & 62.8 \\
  \hline
 \end{tabular*}
 \label{tab}
\end{table} 
\indent The optical depth is the probability at a given time for a source star to be amplified through microlensing by more than a factor of 1.34. The optical depth can be understood as the fraction of solid angle covered by Einstein disks between the observer and the source star (Paczy\'nski \cite{pac}). For a star located at D$_{\rm{s}}$ it is given by
\begin{equation} \label{tds}
 \tau(\rm{D_{s}})=\int_{0}^{\rm{D_{s}}}\frac{\rho_{l}(\rm D_{l})}{\rm{M}} \ \pi \ \rm{R_{E}}^{2} \ \rm{dD_{l}}, \ \ \rho_{l}=\rho_{l}^{disk}+\rho_{l}^{bar}
\end{equation}
where 
\begin{equation}
 \rm{R}_{\rm{E}}=\sqrt{\frac{4G\rm{M}}{c^{2}}\,\frac{\rm{D_{l}(D_{s}-D_{l})}}{\rm{D_{s}}}} 
\end{equation}
is the Einstein radius, D$_{\rm{l}}$ is the distance of the lens and $\rho_{l}$ is the lens density. If the sources are spread along the line of sight, one has to average over all distances and luminosities:
\begin{equation}
 \tau=\frac{\int_{0}^{\infty}\int_{L_{min}({\rm D_{s}})}^{L_{max}} \, \tau({\rm D_{s}) \, D_{s}}^{2} \, \Phi(L) \, {\rm dD_{s}} \, {\rm d}L}{\int_{0}^{\infty}\int_{L_{min}({\rm D_{s}})}^{L_{max}} \, {\rm D_{s}}^{2} \, \Phi(L) \, {\rm dD_{s}} \, {\rm d}L}.
\end{equation} 
\begin{figure}
 \resizebox{\hsize}{!}{\includegraphics{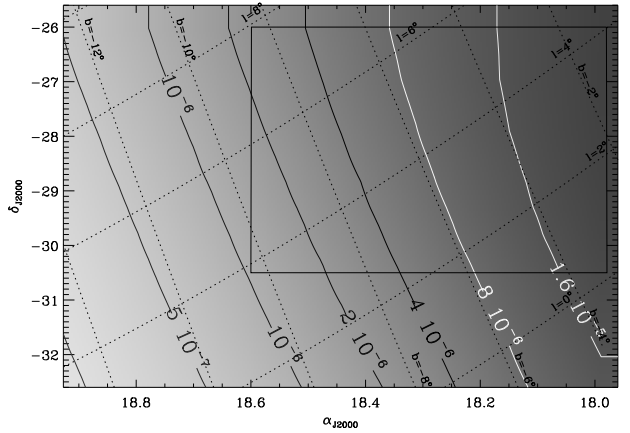}}
 \caption{Optical depth for sources located at the distance of Sgr (24 kpc). The box corresponds to the approximate location of the MACHO fields.}
 \label{tausgr}
\end{figure}
Fig.\ref{tausgr} shows the optical depth map for sources located in Sgr ($\tau_{Sgr}$). This optical depth varies between $\sim$10$^{-6}$ and $\sim3\times 10^{-5}$ in the MACHO fields. This is a factor $\sim$6-7 higher than for a Galactic source, as shown in Table \ref{tab} for three directions and two limiting magnitudes (V=21 mag and V=23 mag).
The first direction (l,b)=(2.7\de,-3.3\de) corresponds to the mean MACHO field published in A00, the second direction (l,b)=(5.0\de,-5.5\de) is located on Sgr's main axis, and the third direction at (l,b)=(6.0\de,-9.0\de) corresponds to the lower-most MACHO field (Popowski et al. \cite{pop}). The higher optical depth for a source in Sgr is due to the greater distance ratio which induces an increase of the Einstein radius. 
%
%
\subsection{Event rate}
\indent The event rate for a source star located at D$_{\rm{s}}$ and lensed by a single mass population can be calculated by (Roulet \& Mollerach \cite{rm})
\begin{figure}
 \resizebox{\hsize}{!}{\includegraphics{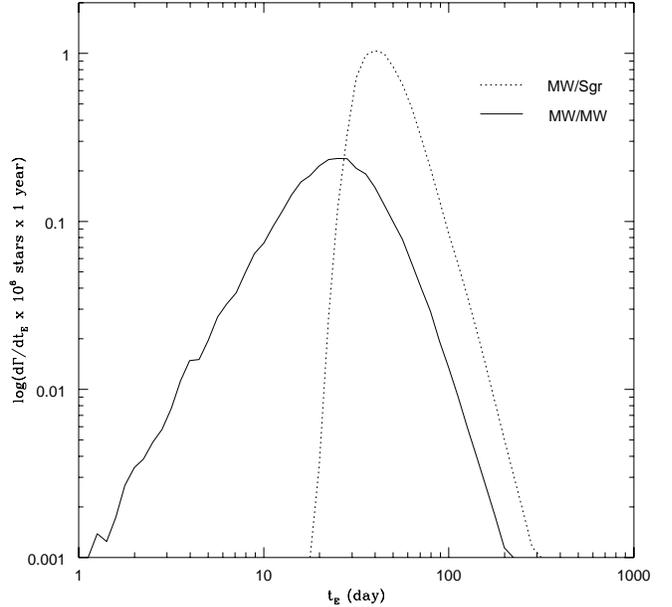}}
 \caption{Event rate as a function of duration of the event calculated for a field centred on (l,b)=(5.0\de,-5.5\de). The mass of the lenses is set to 1 M$\odot$.}
 \label{gam_m0}
\end{figure}
\begin{figure*}
 \resizebox{\hsize}{!}{\includegraphics{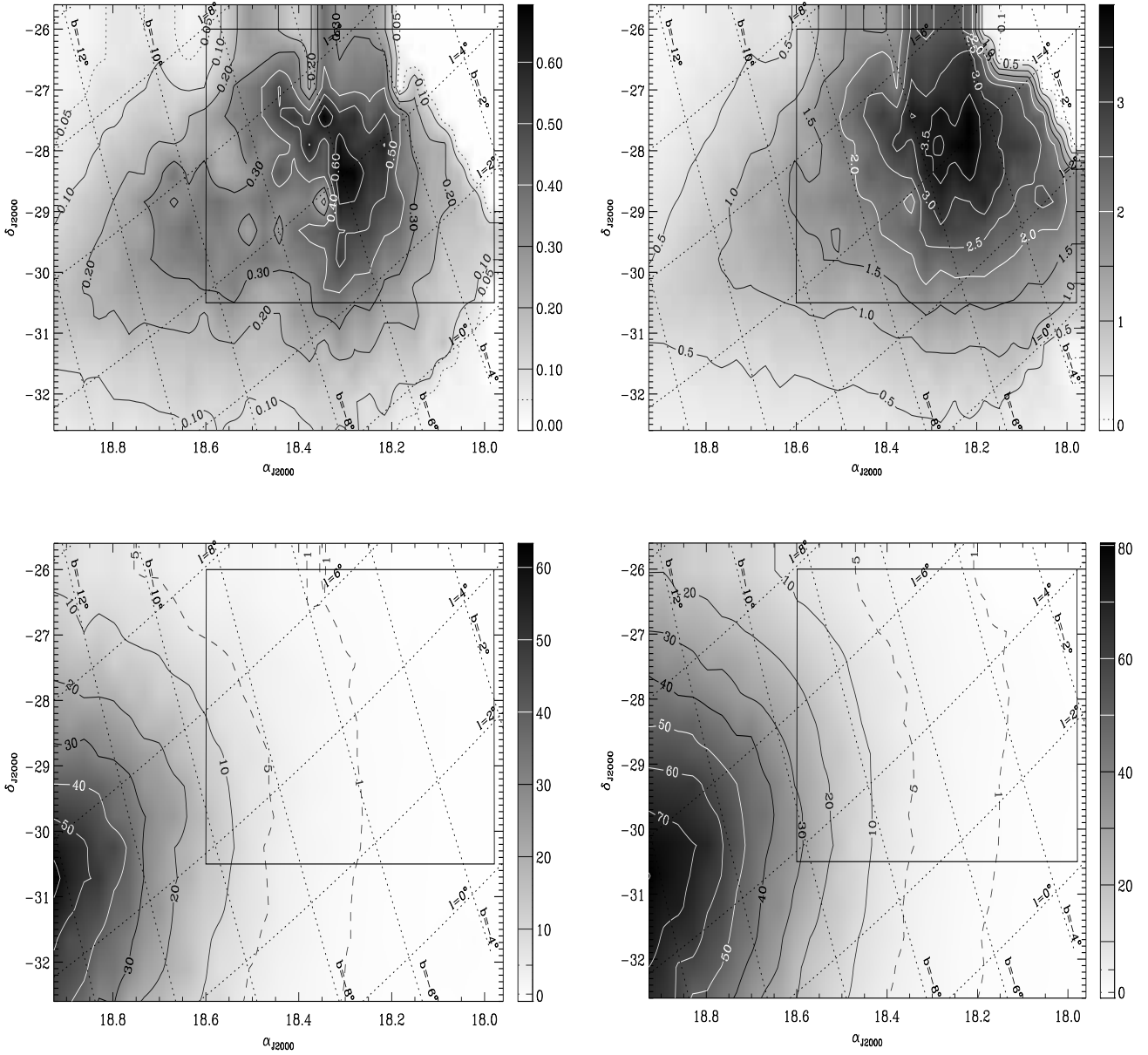}}
 \caption{Upper panels: Number of microlensing events involving a source in Sgr per year and per square degree for a limiting magnitude of respectively V=21.0 mag (left) and V=23.0 mag (right) and assuming a detection efficiency of 100$\%$ for all time-scales. Lower panel: fraction of microlensing events involving a source in Sgr for a limiting magnitude of respectively V=21.0 mag (left) and V=23.0 mag (right). Contours are labelled in percent. The dashed lines are not equidistant with the other contours. The box corresponds to the approximate location of the MACHO fields.}
 \label{maps}
\end{figure*}
\begin{equation}
 \frac{\rm{d}\Gamma}{\rm{d}t_{E}}({\rm D_{s},t_{E}})=\frac{2}{\rm{M}} \ \int_{0}^{2\pi}\int_{0}^{\rm{D_{s}}}\rho_{l}\bigg(\frac{\rm{R_{E}}}{t_{E}}\bigg)^{4}\,f(\mathbf v^{\bot})\,\rm{d}\gamma\,\rm{dD_{l}}
 \label{gameq} 
\end{equation}
where ${\rm t_{E}=R_{E}/|{\mathbf v^{\bot}|}}$ is the Einstein ring crossing time, ${\mathbf v^{\bot}}$ is the lens velocity relative to the line of sight and projected onto the lens plane, f(${\mathbf v^{\bot}}$) is the velocity distribution and $\gamma$ is the position angle of ${\mathbf v^{\bot}}$. 
Fig. \ref{gam_m0} and Table \ref{tab} show the event rate distribution for the ideal case of 100$\%$ detection efficiency for all time-scales and a single mass lens population of 1 M$\odot$. The time-scale of a MW/Sgr event ($\langle\rm{t}_{E}\rangle\sim60$ days) is longer by a factor $\sim$1.3 than the MW/MW events ($\langle\rm{t}_{E}\rangle\sim45$ days). These time-scales increase for both MW/MW and MW/Sgr events towards higher $|b|$ because of the increasing contribution of the Disk lenses located between the Sun and the Galactic Centre. The time-scale dispersion is slightly lower for MW/Sgr events because the velocity dispersion of the source is negligible in this case.
\subsection{Fraction of events involving Sgr sources}
To estimate the number of events involving sources in Sgr we convolved the event rate distribution with a Salpeter mass function $\xi{{\rm (M)}}\varpropto \rm{M}^{-2.35}$ with an upper cut-off at 1 M$\odot$ and a lower cut-off at the brown dwarf limit (0.08 M$\odot$). The total number of events for a source located at D$_{s}$ is (Roulet \& Mollerach \cite{rm}):
\begin{equation}
 \Gamma(D_{s})=\frac{\int_{0}^{\infty}\int_{\rm {0.08\,M_{\odot}}}^{\rm{1\,M_{\odot}}}\frac{\rm{M_{\odot}}}{\rm{M}}\,\gamma(\rm{D_{s},t_{E},M})\,\xi(\rm{M})\,\rm{dt_{E}}\,\rm{dM}}{\int_{\rm {0.08\,M_{\odot}}}^{\rm{1\,M_{\odot}}}\xi(\rm{M})\,\rm{dM}}
 \label{gam_ds}
\end{equation}
where
\begin{equation}
 \gamma(\rm{D_{s},t_{E},M})=\frac{\rm{d}\Gamma}{\rm{d}t_{E}}\bigg({\rm D_{s},t_{E}\sqrt{\frac{\rm{M_{\odot}}}{\rm{M}}}}\bigg).
\end{equation}
Similar to the optical depth for sources spread along the line of sight, one has to integrate Eq.\ref{gam_ds} over the range of source distances and luminosities. The upper panels of Fig.\ref{maps} show the maps of the number of events expected per year and per square degree ($\Sigma_{\rm Sgr} \times \Gamma_{\rm Sgr}$) originating from sources in Sgr for two limiting magnitudes (V=21 mag and V=23 mag). These maps show that the highest rate is expected near (l,b)=(4\de,-5.5\de). Closer to the Galactic Plane the rate is limited by the extinction which becomes too important to reach the sources, whereas at higher $|b|$ the optical depth drops too fast to take advantage of the increasing source density in Sgr. It is interesting to note that the region of highest rate is located near the centre of the MACHO fields.\\
\indent The lower panels of Fig.\ref{maps} shows the fraction of MW/Sgr events $f_{Sgr}=(\Sigma_{\rm Sgr} \times \Gamma_{\rm Sgr})/(\Sigma_{\rm tot} \times \Gamma_{\rm tot})$ for V$_{\rm lim}$=21 mag (left) and V$_{\rm lim}$=23 mag (right) respectively. It is obvious from these maps that $f_{Sgr}$ depends on the limiting magnitude of the survey. This is due to the variation of the star ratio ($\Sigma_{\rm Sgr}/\Sigma_{\rm tot}$) with the limiting magnitude (Fig.\ref{lf}). The star ratio reaches a minimum between V=20 mag and V=21 mag (depending on the extinction), corresponding to a magnitude range fainter than the Bulge Turn Off but brighter than the Sgr Turn Off. After this minimum, the star ratio increases continuously towards fainter magnitudes. Thus, the deeper the survey, the higher $f_{Sgr}$. The MW/Sgr events become dominant only near the highest density region of the dwarf galaxy located at (l,b)=(5.5\de,-14.0\de). Note that the rates are calculated for an efficiency of unity whereas the event detection efficiencies generally increase with the event duration. Hence, since MW/Sgr events have a longer time-scale than the MW/MW events, these maps only present lower limits for $f_{Sgr}$ in current experiments.
\subsection{Uncertainties in the MW/Sgr event rates}
\begin{figure}
 \resizebox{\hsize}{!}{\includegraphics{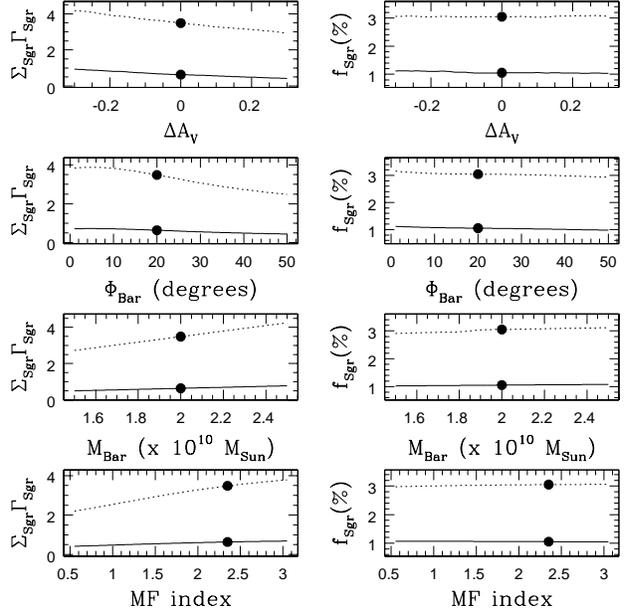}}
 \caption{Variations on the rate (left) and fraction of MW/Sgr events (right) for some parameters of our model (from top to bottom: extinction, Bar angle, Bar mass and mass function index). The full (resp. dotted) line corresponds to a limiting magnitude of V$_{\rm lim}$=21 (resp. V$_{\rm lim}$=23). The filled circle corresponds to the value calculated in our reference model for a field of 1 deg$^{2}$ centred on (l,b)=(4.0$^{\circ}$,-5.5$^{\circ}$).}
 \label{uncertainties}
\end{figure}
\indent Estimating the uncertainties in the MW/Sgr event rate is not an easy task because the parameter space is so wide. Different rates may be obtained for different Bar models, luminosity functions, lens mass functions, extinctions, etc... . In this section we estimate the effect of varying each parameter one by one in our model. This should provide a clear idea of the uncertainties. To do this, we will focus on the direction (l,b)=(4.0$^{\circ}$,-5.5$^{\circ}$), close to the location where the highest MW/Sgr rate is expected.
\subsubsection{Source density in Sgr}
\indent  The uncertainty in the source density in Sgr is relatively high ($\sim$40$\%$) as the transformation of RR Lyrae into stars is based on only five variables. However, correcting the MW/Sgr rate is straightforward since the event rate scales linearly with the surface density in Sgr. A different source density in Sgr would influence only the rate but not the spatial distribution of the events. This is also true for $f_{Sgr}$ at $|b|\lesssim$10$^{\circ}$, where the MW/MW events dominate.\\
\indent Another uncertainty may result from the luminosity function of Sgr. As stated above, we had to extrapolate this function for magnitudes fainter than M$_{\rm V}\sim$4.5 because of completeness problems. This magnitude corresponds to an apparent magnitude of V$\sim$22.5-23.5 mag (depending on extinction). Hence the uncertainty in the luminosity function of Sgr has no effect on our results for V$_{\rm lim}$=21 and should only mildly influence the results for  V$_{\rm lim}$=23.
\subsubsection{Extinction}
\indent The extinction determines how faint one can reach the luminosity function of the sources. Fig. \ref{uncertainties} (upper panels) shows how the MW/Sgr event rate and fraction evolve if we let the extinction vary up to $\pm$0.3 mag about the assumed value. As expected, the rate increases when extinction drops. The error on the rate is $\sim$10$\%$ for a deviation of $\pm$0.2 mag about A$_{\rm V}$. The fraction of events is only marginally influenced by uncertainties in A$_{\rm V}$ because extinction affects the MW/MW and MW/Sgr rates in an almost similar way. 
\subsubsection{Lens mass function}
\indent In our calculation we used a simple Salpeter function for the mass distribution of the lenses. While this is a fair approximation in most cases, we investigate the effect of changing the exponent of the mass function and let it vary between 0.55 and 3.05:
\begin{equation}
 \xi(\rm M) \varpropto {\rm M}^{-\alpha}, \ \ \ 0.55<\alpha<3.05.
\end{equation}
 Fig.\ref{uncertainties} shows that flat mass functions induce lower rates. For instance, an index $\alpha$=1.35 would produce $\sim$20$\%$ less events. A flat mass function at the low mass end seems to be favored by the luminosity function towards the Bulge (Holtzman et al. \cite{holtz}, Zoccali et al. \cite{zoc}). However there remains uncertainties in the binaries fraction, and also a flat mass function would be in conflict with the distribution of time-scales of observed microlensing events towards the Galactic Centre. The fraction of MW/Sgr events is almost independent of the mass function.
\begin{figure}
 \resizebox{\hsize}{!}{\includegraphics{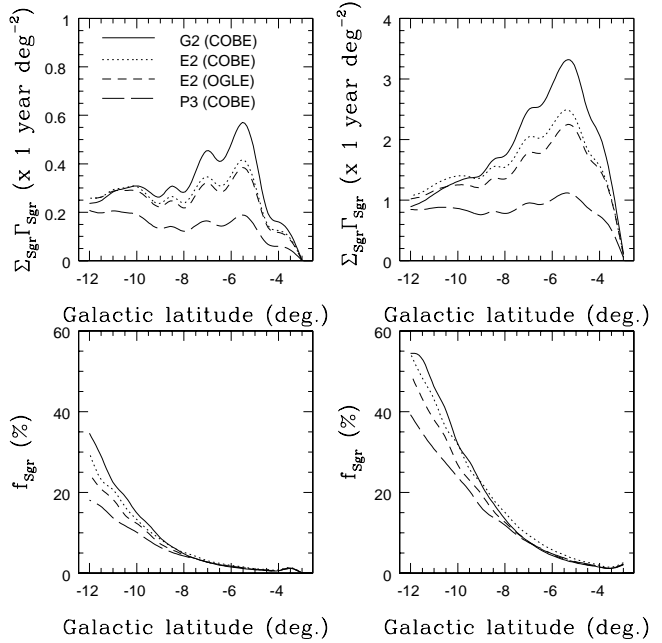}}
 \caption{{\bf Upper panels:} Cross-section at l=5$^{\circ}$ of the MW/Sgr event rate for different Bar models (solid line: G2, dotted line: E2(COBE), short dash: E2(OGLE) and long dash: P3(COBE)). The left (resp. right) panel correspond to a limiting magnitude of V$_{\rm lim}$=21 (resp. V$_{\rm lim}$=23). All the models are normalized to a Bar mass of 2.0 10$^{10}$M$_{\odot}$. {\bf Lower panels:} Same as above for the fraction of MW/Sgr events.}
 \label{models}
\end{figure}
\subsubsection{Bar model}
\indent The Galactic model is probably the most important parameter in our study. We will discuss here only the consequence of modifying the Bar since the majority of events are believed to be caused by lenses located within this component. Dwek et al. (\cite{dwek}) fitted three families of analytical functions (Gaussian-type, exponential-type and power-law-type) to the infra-red surface brightness measured by the COBE/DIRBE instrument towards the Galactic Centre. We took as a `reference' model the one corresponding to their best fit (which they named G2). However, other models fit the COBE/DIRBE map almost equally well. We consider now two other models:
\begin{equation}
  \rho_{\rm E2} = 25.65 \times  \rm{exp}[-r] \ \ \ (\rm{M}_{\odot}.\rm{pc}^{-3})
\end{equation}
and
\begin{equation}
  \rho_{\rm P3} = 39.16 \times  \bigg(\frac{1}{1+r^{2}}\bigg)^{2}  \ \ \ (\rm{M}_{\odot}.\rm{pc}^{-3})
\end{equation}
where 
\begin{equation}
  r= \bigg[\bigg(\frac{x'}{x_{0}}\bigg)^{2}+\bigg(\frac{y'}{y_{0}}\bigg)^{2}+\bigg(\frac{z'}{z_{0}}\bigg)^{2}\bigg]
\end{equation}
x', y' and z' are defined in the same way as Eq.\ref{s4}. The scale lengths are ($x_{0},y_{0},z_{0}$)=(0.74,0.16,0.27)kpc and (0.90,0.23,0.28)kpc for $\rho_{E2}$ and $\rho_{P3}$ respectively. The angle between the x-axis of the Bar and the Sun-GC direction are 41.3$^{\circ}$ (E2 model) and 35.4$^{\circ}$ (P3 model). Additionally, we consider model E2 with different parameters: ($x_{0},y_{0},z_{0}$)=(0.90,0.385,0.25)kpc and an angle of 24$^{\circ}$. These latter parameters correspond to the best fit of the OGLE red clump star distribution towards the Galactic Centre (Stanek et al. \cite{stanek}). All the above Bar models are normalized in order to get a Bar mass of 2$\times$10$^{10}$M$_{\odot}$. \\
\indent We use these models to calculate the MW/Sgr rate and $f_{Sgr}$ at Galactic longitude l=5$^{\circ}$ for -12$^{\circ}<b<$-3$^{\circ}$, following approximately Sgr's main axis. These rates are compared to our reference model in Fig.\ref{models}. The successive bumps in the rates are due to the variation of extinction along the strip. All the models produce a maximum rate near b$\sim$-5.5$^{\circ}$. The rapid decrease of the rates at lower latitudes are due to the increasing extinction where Sgr sources can no longer be detected. Similar to the Bulge self-lensing case, the G2-model is the most efficient for producing MW/Sgr events while the P3-model is the least efficient. Surprisingly, both E2 models (COBE and OGLE) seem to produce the same amount of MW/Sgr events in spite of their different parameters. One would have expected a higher rate for the OGLE Bar model because of its higher inclination. For instance, in Fig. \ref{uncertainties}  we show how the event rate evolves if we let the angle of our reference model vary, confirming that a lower angle favors a higher rate. However, the OGLE model has longer scale-lengths, inducing a lower density (remember all Bar models are normalized to the same mass), compensating the effect of the lower Bar angle. Of course there is a strong dependence between the Bar mass and the MW/Sgr rate. In Fig.\ref{uncertainties} we let the Bar mass vary between 1.5$\times 10^{10}$ and 2.5$\times 10^{10}$M$_{\odot}$, encompassing the estimation range of different authors (e.g. Gerhard \& Vietri \cite{gv}, Kent \cite{kent}, Zhao \cite{zbar}). Indeed, the MW/Sgr rate scales almost linearly with the Bar mass with a slight non-linearity caused by the Disk. Finally, the fraction of MW/Sgr events is almost independent of the model, showing that the efficiencies for a Bar model to produce MW/MW and MW/Sgr events are similar.
\section{Application to observations}
\subsection{MACHO}
\begin{figure}
 \resizebox{\hsize}{!}{\includegraphics{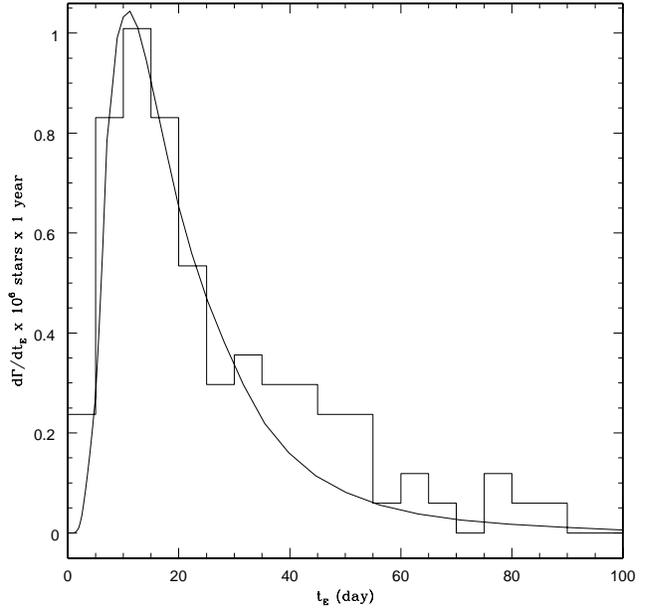}}
 \caption{Time-scale distribution of microlensing events for a field centred on (l,b)=(2.7\de,-3.3\de), corresponding to the average location of the MACHO fields (A00). The limiting magnitude is set to V=23 and the extinction is A$_{\rm V}$=1.8 mag. The detection efficiency is taken from A00. The distribution resulting from our reference model is compared to the observations of the MACHO team (Alcock et al. \cite{macho2}, A00), normalized to an exposure of 10$^{6}$ stars $\times$ 1 year (histogram).}
 \label{gam}
\end{figure}

\indent Recently the MACHO collaboration reported 99 events towards the Galactic Centre (Alcock et al. \cite{macho2}; A00). Their detection is based on the Difference Image Analysis technique (DIA, Alcock et al. \cite{macho_dia}), which is sensitive to microlensing events on stars as faint as V=23 mag, provided the peak magnification of these stars is brighter than V$\sim$21 mag. We computed the expected event rate for their average field (l=2.7\de,b=-3.3\de) using the detection efficiency published in A00. The resulting event rate distribution is compared to the observations (normalized to an exposure of 10$^{6}$ stars $\times$ 1 year) in Fig.\ref{gam}, showing the relatively good match of our reference model to the observed event distribution, especially at short time-scales. We find that the fraction of MW/Sgr events is 1.1$\%$, thus the number of MW/Sgr events published in A00 is of the order of unity. \\
\indent The MACHO collaboration has now collected seven seasons of data in 94 fields ($\sim$45 deg$^{2}$) located between b$\simeq$-2\de and b$\simeq$-9\de (Popowski et al. \cite{pop}). When these data will have been processed with the DIA technique, there will be potentially $\gtrsim$1\,000 events available and some of these will be located at higher $|b|$ where the probability for MW/Sgr events increases. It is thus very likely that some of these events will involve sources in Sgr.
\subsection{OGLE}
\indent The OGLE collaboration published the largest catalogue to date, comprising a set of 214 microlensing events to a limiting magnitude of V$\sim$21 mag towards the Galactic Centre (Udalski et al. \cite{ogle2}). Their detection process is based on PSF photometry so that one has to take into account the blended events to calculate the rate (Alard \cite{a97}). These events are caused by magnification (through microlensing) of unresolved sources located inside the seeing disk of a resolved star to which the event is attributed. This causes the detected stars to act as multiple sources (Han \cite{han}). A blended event mimics a classical microlensing event with a reduced Einstein radius (Di Stefano \& Esin \cite{dise}). Thus, the rate associated with these blended events is calculated in the same manner as in Eq.\ref{gameq}, except that one has to take into account the reduction of the Einstein radius by a factor 
\begin{equation}
 \rm{r_{b}}=\sqrt{2\Bigg(\frac{A_{b}}{\sqrt{A_{b}^{2}-1}}-1\Bigg)}
\end{equation}
with
${\rm A_{b}}=0.34\,(l_{b}/L)\,(\rm{D_{s}/10\,pc})^{2}+1.34$
where $L$ is the absolute luminosity of the source being micro-lensed and $l_{b}$ is the apparent luminosity of the blending star. A good estimation of the apparent magnitude distribution towards the Galactic Centre is provided by a power law $\varpropto l_{b}^{-2}$ (Zhao et al. \cite{zsr}). Furthermore, one has to normalize this rate by the fraction of the field covered by all the seeing disks associated with the detected stars because only events on sources located into the seeing disk of the resolved stars are detected. For OGLE, this ratio is $\sim$0.5 (Alard \cite{a97}). We take the detection efficiency from Udalski et al. (\cite{ogle1}). It results that the fraction of events due to microlensing of Sgr sources in OGLE is $\sim$0.7$\%$ at (l,b)=(2.7\de,-3.3\de). This value is slightly lower than for MACHO because sources fainter than V$\sim$21 mag contribute only through blended events which have lower time-scales than resolved events and are thus more difficult to detect. Furthermore, the OGLE fields are spread over a wide range in galactic longitude (-10\de$\lesssim$l$\lesssim$10\de) and some of these fields may not contain any stars of Sgr. The number of events involving a source in Sgr in the OGLE catalog is thus also of the order of unity. 
%
%
\section{How to detect MW/Sgr events and why}
\begin{figure}
 \resizebox{\hsize}{!}{\includegraphics{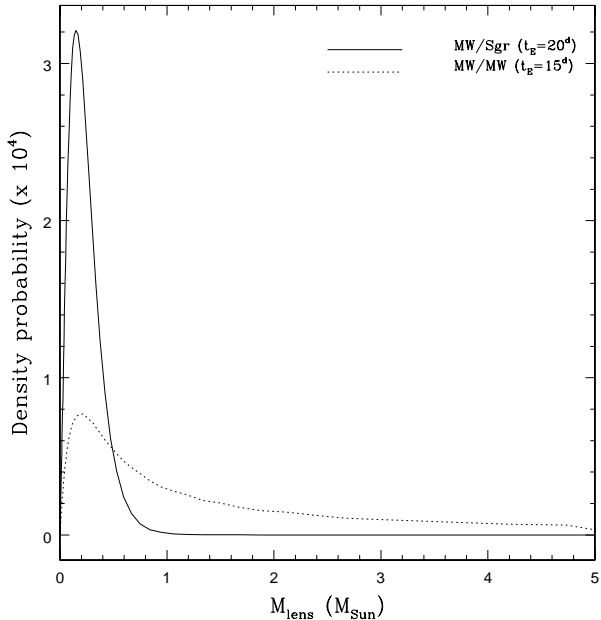}}
 \caption{Normalized density probabilities of the lens mass. The solid line corresponds to a MW/Sgr event with t$_{\rm E}$=20$^{\rm d}$ whereas the dotted line corresponds to a MW/MW event with a time-scale of 15 days.}
 \label{mlprob}
\end{figure}
To detect MW/Sgr events, one needs to reach Sgr's turn off. This is only feasible for fields with not a sufficiently low extinction (A$_{\rm V}\lesssim$2 mag) and requires sensitivity to magnitudes fainter than $\sim$21 mag. Fortunately, image subtraction methods are improving rapidly (Alard \& Lupton \cite{al}, Alcock et al. \cite{macho_dia}, Alard \cite{a_isis}) and allow us to reach magnitudes down to V$\sim$23 because (1) crowding is virtually suppressed, (2) the micro-lensed star need not be pre-registered in a catalog and can be detected at any time provided the peak magnification makes it brighter than the detection limit. Detection of MW/Sgr events is thus already within the reach of current microlensing surveys. One needs, however, a criterion to discriminate between MW/MW and MW/Sgr events. The heliocentric radial velocity of a Sgr member star is $\sim$140$\pm$10 km.s$^{-1}$. This is sufficiently large to allow an unambiguous separation between Sgr and Bulge sources (see for instance Fig. 4 of Ibata et al. \cite{igi}). MW/Sgr events can thus easily be spotted with a single spectroscopic measure which can be performed well after the event occured. \\
\indent There are several reasons why detection of MW/Sgr events is interesting. First, they provide stronger constraints on the lens mass than MW/MW events. Determining the lens mass for a single lens microlensing event is a highly degenerate problem because the mass depends simultaneously on the proper motions and distances of both the lens and the source (through the event time-scale). Since the distance ($\sim$24 kpc)  and proper motion (2.1 mas.year$^{-1}\simeq$250 km.s$^{-1}$ towards the Galactic Plane) of Sgr are relatively well constrained (Ibata et al. \cite{iwgis}), the lens mass determination becomes much more simple. For example, we present in Fig. \ref{mlprob} the lens mass density probabilities for MW/Sgr and  MW/MW events with time-scales of 20 days and 15 days respectively. These curves have been constructed by simulating 10$^{6}$ lens/source configurations drawn from the values and uncertainties quoted in Sect. 2. One sees that the density probability of MW/MW event is skewed towards large masses and that it can result from a broad range of lens masses. On the other hand, the lens mass of a MW/Sgr event is much better constrained and it is possible for a \emph{single} event to associate an upper limit to the lens mass. Of course, the  MW/MW sample is large and allows a statistical estimation of the lens mass population, but something like $\sim$10-20 MW/Sgr events could provide additional constraints on this determination.\\
\indent Another interesting point is the longer time-scales of MW/Sgr events relative to MW/MW events. They should therefore be easier to detect by microlensing surveys. This is important because the lens mass scales as the square of the event time-scale. It results that MW/Sgr events probe the lens masses spectrum deeper into the low mass end than MW/MW events. In our reference model we found t$_{\rm E}$(MW/Sgr)$\simeq$1.3 t$_{\rm E}$(MW/MW), resulting in a low mass detection limit $\sim$1.7 lower for a Sgr source than for a Galactic source. \\
\indent In this paper we investigated only the formal case where a source in Sgr is micro-lensed by a deflector in the Disk or in the Bulge. Other configurations involving a source in Sgr are also possible. Lines of sight towards Sgr's northern stream probe regions located \emph{behind} the Bulge and can be used to put constraints on lens populations in these regions. Potential candidates for such populations are a warp and/or flare of the Disk or a yet-undetected stream of stars. Also, while lines of sight towards the Magellanic clouds probe the outer Halo, Sgr could be used as a target to put constraints on the dark mass content of the inner Halo. We showed that the fraction of MW/Sgr events was almost independent of the Bar model. A significantly higher fraction of MW/Sgr events would hint at an additional intervening lens population.
\section{Conclusion}
\indent We have shown that some of the microlensing events towards the Galactic Centre may be due to magnification of sources in Sgr. These events have a longer time-scale than the MW/MW events and occur preferentially on faint sources, below  Sgr's turn off at V$\gtrsim$21 mag. These events do not contribute significantly in the catalogs published up to now by MACHO and OGLE because these concerned only fields at low $|b|$ where the fraction of MW/Sgr events is $\lesssim$1$\%$. However when one considers a field farther away from the Galactic Plane the contribution of Sgr in the event rate has to be taken into account. The MACHO collaboration has now collected seven seasons of data down to b$\sim$-9\de and some of the new events will plausibly involve sources in Sgr. We caution that any optical depth map inferred from observations might become biased by the presence of Sgr at high $|b|$. \\
\indent A possible way to detect the MW/Sgr events is to perform spectroscopic measurements on all the sources in order to determine their radial velocity. The detection of this kind of events is interesting as it will considerably reduce the degeneracy in the event parameters because the distance and proper motion of the source will be known, allowing us to put strong constraints on the lens mass. Additionally, the Northern stream of Sgr can be used as a microlensing target to probe the potential lens populations located behind the Bulge.\\
%
%
\begin{acknowledgements}
 We thank the anonymous referee for his/her useful advice which helped to improve the paper.
\end{acknowledgements}

\end{document}